\documentclass[preprint2]{aastex}

\usepackage{txfonts}
\usepackage{longtable,lscape,rotating}
\usepackage{multirow}
\usepackage{array}
\usepackage{color}

\newcommand{\ovi}{O{\small ~VI}}                 
\newcommand{\ovii}{O{\small ~VII}}               
\newcommand{\oviii}{O{\small ~VIII}}             
\newcommand{\civ}{C{\small ~IV}}                 
\newcommand{\cvi}{C{\small ~VI}}                 
\newcommand{\siiv}{Si{\small ~IV}}               
\newcommand{\nv}{N{\small ~VI}}                  
\newcommand{\rosat}{{\it ROSAT}}                 
\newcommand{\oqkev}{$1\over4$~keV}               

\shorttitle{SWCX emission in the \oqkev\ band}
\shortauthors{Koutroumpa et al.}

\begin{document}


\title{The solar wind charge-transfer X-ray 
emission in the \oqkev\ energy range: inferences 
on Local Bubble hot gas at low Z}


\author{D. Koutroumpa\altaffilmark{1} and R. Lallement\altaffilmark{1}}
\affil{IPSL/Service d'A\'eronomie - CNRS, Verri\`eres le Buisson, France}
\author{J. C. Raymond\altaffilmark{2} and V. Kharchenko\altaffilmark{2}}
\affil{Harvard - Smithsonian Center for Astrophysics, Cambridge, MA}




\altaffiltext{1}{UMR 7620, IPSL/Service 
d'A\'eronomie, CNRS, Universit\'e de Versailles - Saint 
Quentin en Yvellines, Universit\'e Pierre et Marie Curie,
Verri\`eres le Buisson, 91371, France}
\altaffiltext{2}{Harvard - Smithsonian Center for 
Astrophysics, 60 Garden Street, Cambridge, MA 
02138, US}


\begin{abstract}
We present calculations of the heliospheric solar wind charge exchange (SWCX) emission spectra 
and the resulting contributions of this diffuse background in the \rosat\ \oqkev\ bands. 
We compare our results with the soft X-ray diffuse background (SXRB) emission detected in 
front of 378 identified shadowing regions during the \rosat\ All-Sky Survey \citep[][]{snowden00}. 
This foreground component is principally attributed to the hot gas of the so-called Local Bubble 
(LB), an irregularly shaped cavity of $\sim$ 50-150 pc around the Sun, which is supposed to contain $\sim$10$^6$ K plasma. 
Our results suggest that the SWCX emission from the heliosphere is bright enough to account for 
most of the foreground emission towards the majority of low galactic latitude directions, 
where the LB is the least extended. On the other hand, in a large part of directions with galactic 
latitude above 30 degrees the heliospheric SWCX intensity is significantly smaller than the measured one.
However, the SWCX R2/R1 band ratio differs slightly from the data 
in the galactic center direction, and more significantly in the galactic anti-centre 
direction where the observed ratio is the smallest. Assuming that both SWCX and hot gas emission 
are present and their relative contributions vary with direction, we tested a series of thermal 
plasma spectra for temperatures ranging from 10$^{\,5}$ to 10$^{\,6.5}$ K and searched for a 
combination of SWCX spectra and thermal emission matching the observed intensities and band ratios, while simultaneously 
being compatible with \ovi\ emission measurements. In the frame of collisional equilibrium models and for solar 
abundances, the range we derive for hot gas temperature and emission measure cannot reproduce the 
Wisconsin C/B band ratio. This implies that accounting for SWCX contamination does not remove these 
known disagreements between data and classical hot gas models. We emphasize the need for additional atomic data, 
describing consistently EUV and X-ray photon spectra of the charge-exchange emission of heavier solar wind ions.
\end{abstract}





\section{Introduction}
The diffuse soft X-ray background (SXRB), first observed in the 70's \citep{bowyer68,williamson74,sanders77} has since 
been shown to be the sum of local and distant sources. Above 2 keV it is dominated by the extra-galactic 
background, itself a combination of unresolved point sources and warm-hot interstellar medium (WHIM) diffuse 
emission \citep{hasinger93}. At lower energies it is dominated by the galactic halo 
\citep{burrows91,snowden94a}, and finally below 0.3 keV it is mainly due
to the unabsorbed emission from hot gas filling the so called Local Bubble (LB) 
\citep{mccammon83,bloch86,snowden90a,snowden90b}, a cavity devoid of dense gas extended
at high latitudes and connected to the halo \citep{frisch83,welsh98,lallement03}. The main 
tools used to disentangle local and distant emission are the `shadowing' experiments, i.e. spatial variations of intensity and spectral 
characteristics around and towards dense, soft X-ray absorbing clouds \citep[e.g.][]{herbstmeier95}. \cite{snowden98,snowden00} 
used more than 370 \rosat\ shadows to produce almost full-sky mapping of the `unabsorbed' component of the emission, i.e. the LB contribution.

This was the generally accepted scenario until the discovery of X-ray emission in comets 
\citep[][]{lisse96} and the identification of the emission mechanism as charge-exchange (CX) reactions between the 
highly charged heavy solar wind ions and the cometary neutrals \citep[][]{cravens97}. \cite{cox98} suggested 
that the CX reactions should also occur between heavy SW ions and interstellar neutrals (H and He) in interplanetary space and that 
the resulting X-ray emission (SWCX) should have an impact on the SXRB interpretation. \cite{cravens00} estimated that the quiescent 
level of SWCX emission could be of the same order as the SXRB component attributed to the LB.

This interplanetary, heliospheric emission is time-dependent because 
of the intrinsic variable nature of the solar wind. Short time-scale variations tend to be 
washed out by integration along the line-of-sight and the size of the emitting region \citep{cravens01}, 
but longer term variations, including those related to the solar cycle, can cause more persistent changes 
in the heliospheric emission level. In addition, there is a contribution from the Earth's magnetosphere, due 
to charge-transfer with exospheric neutrals. Such emission, studied in detail by \cite{robertson06}, reacts 
instantaneously to solar wind variations and magnetosphere shape variations, leading to a high variability and the occurence 
of high intensity peaks following solar events. The spectral characteristics of some spectacular enhancements have been 
recently recorded by the XMM and Suzaku satellites \citep[][]{SCK04,henley08}. For such events both magnetospheric and 
heliospheric contributions may be present. 

Very likely most of the sharp increases of terrestrial origin have been removed from the 
\rosat\ map along with the cleaning procedure of the Long Term Enhancements \citep[LTEs][]{snowden94b}, 
as well as some heliospheric increases, especially towards the downwind side of the interstellar flow 
where the gravitational cone of focused helium is the most reactive region. Indeed, most points 
on the sky in the \rosat\ map were observed several times over the course of at least two days, 
allowing identification and removal of periods of enhanced emission. A debate is still maintained, 
though, about the actual level of the quasi-stationnary heliospheric contribution to the \rosat\ maps 
of unabsorbed emission. This contribution is extremely difficult to detect from time variations. 
On the other hand, SCWX and hot gas thermal emission have different spectral properties, 
i.e. the observed spectral information should help to disentangle the two processes. This is 
the subject of the present study. For a recent review of all types of SWCX phenomena see \cite{bhardwaj07}.

The first estimates of the stationary heliospheric contribution \citep{cravens00,lall04CX} were based on simplifying assumptions
about the spectral characteristics of the SWCX emission. Since then the existence of the SWCX phenomenon has motivated 
theoretical work on exact photon yield values for the charge transfer collisions \citep{kharchenko00,pepino04}, as well as a number 
of laboratory experiments devoted to the CX emission mechanism. For a recent review see \cite{wargelin08}.

The spatial distribution of magnetospheric and heliospheric SWCX emission 
was modeled by \cite{robertson03}, revealing significant variations in 
brightness as a function of earth location, line-of-sight direction, 
and activity phase. \cite{kout06} computed similar maps for a few 
specific energy bands after including \cite{pepino04}'s detailed CX emission 
spectra for C, N, O, and Ne ions. \cite{lall04CX}, taking into account 
the specific viewing geometry of \rosat\, showed that the heliospheric background 
in the \oqkev\ band was nearly isotropic and could have been unnoticed in the 
All-Sky Survey maps, while accounting for a large portion of the signal, 
and possibley the major part at low galactic latitudes.

Using both the stationnary and time-dependent models \cite{kout07} modelled four high-latitude 
shadowing observations and showed that in the 3/4 keV band, where the oxygen lines 
(\ovii\ triplet at 0.57 keV and \oviii\ line at 0.65 keV) are dominant, 
the SWCX emission from the heliosphere can account for all the unabsorbed, 
local component of the SXRB, with no need of a LB emission. In parallel, 
the solar wind contribution to the background and its variability have been shown 
to be responsible for some discrepant measurements \citep{smith07} and for supposedly 
low-energy counterparts of distant objects \citep{bregman06}.

A 10$^6$ K plasma, however, has very little emission in the 3/4 keV band and mainly emits 
in the \oqkev\ band. The \cite{kout07} results, while not requiring any LB emission, 
therefore do not preclude the existence of 10$^6$ K gas. Exact calculations of the SWCX spectra 
and intensities below 0.3 keV are mandatory if one wants to disentangle LB hot gas diffuse emission from the SWCX background. 
In this paper we examine the SWCX contribution to the \oqkev\ spectral region, compare this contribution to observations 
and also to contributions from hot gas at different temperatures.

Independently of the SWCX contribution, a number of results have somewhat contradicted the 
interpretation of the unabsorbed soft X-ray background as the LB 10$^6$ K gas emission.

i) Data from the NASA EUVE satellite and from the dedicated CHIPS mission did not detect the EUV 
emission expected from surrounding 10$^6$ K gas \citep{jelinsky95,hurwitz05}. It has been 
suggested that a very low metal abundance may be responsible for this non-detection, but the 
required depletion level corresponds to the physical state of very dense clouds, which is 
unlikely for 10$^6$ K, tenuous gas. 

ii) The pressure of this hot gas derived from the X-ray background is far above the pressure within 
the local interstellar cloud and other clouds embedded in the LB \citep{lallement98,jenkins02}.

iii) Low latitude absorption measurements of highly charged ions such as \siiv\, \civ\ and \ovi\ formed in conductive 
interfaces between the hot (10$^6$ K) gas and embedded cold:warm clouds do not seem to correspond to expectations 
from the models \citep[][]{slavin02,indebetouw04}. Column densities of \siiv\ and \civ\ are too small and line-widths 
too narrow \citep{welsh05}, and \ovi\ is detected only at the periphery of the Local Cavity, while one would also 
expect interfaces between the hot gas and the local clouds \citep[][]{welsh08}.

iv) Fundamental discrepancies arise also when comparing the Wisconsin sounding rocket survey data 
in the B and C bands, and the \rosat\ All-Sky Survey data in the R1 and R2 bands. The four bands are pictured 
in figure \ref{fig:spec1_4}-\textit{upper panel}. In the low energy (0.1-0.2 keV) B band \citep{bloch86,snowden94a}, the intensity seems 
to be higher than what is predicted by thermal emission models. This has been particularly well 
demonstrated by \cite{bellm05} who have made a global study over the 0.1-0.3keV interval. 
According to this work, a best fit to all energy bands is provided by very low metallicity gas at 
10$^{5.85}$ K, but inspection of their results (see their Figure 4) reveals significant discrepancies 
between measured and observed ratios for this best fit solution. Especially, the B/R12 band 
ratio favours a low temperature $\simeq$10$^{5.8}$ K (the B band intensity is high and favours a 
shift of the spectrum towards low energies), while the R2/R1 ratio favours temperatures above 
10$^6$ K (R2 is relatively high, favouring a shift towards high energies).

Whether or not the existence of the SWCX background can help to explain part or all these 
contradictions is a question that has now to be addressed. This work is a first step in this 
direction. In section \ref{sec:model} we describe the the SWCX emission and spectral model we have 
developed and how we make use in our analysis of the Raymond \& Smith (R-S) hot plasma model. In 
section \ref{CX_IoRosat} we compute the expected SWCX emission and the contribution 
in \rosat\ R1 and R2 bands for each of the 378 shadow regions observed by \cite{snowden00}. 
We compare the SWCX R1+R2 intensity with the unabsorbed component derived by the 
\cite{snowden00} shadow analysis and discuss the distribution of the discrepancies between 
data and the SWCX model. In section \ref{band_ratios} we compute the SWCX model R2/R1 and B/C band 
ratios, as well as the corresponding ratios for hot gas (R-S model) in collisional equilibrium
within a large temperature range. We compare the modeled ratios with the observed band ratios 
during the two (Wisconsin and \rosat ) surveys. In section \ref{combination} we search for a 
combination of SWCX and hot gas emission compatible with the observed intensities and band ratios and 
we compare those solutions with observational constraints from \ovi\ and EUV background measurements. In 
section \ref{discussion} we discuss the results and draw some conclusions.

\section{Model description}
\label{sec:model}

\subsection{SWCX Model}
\label{sec:CXmodel}

The basic model calculating the SWCX emission in the inner heliosphere was thoroughly presented in 
\citet{kout06}; parameters appropriate for the \rosat\ All-Sky Survey are discussed in \ref{CX_IoRosat}. 
We calculate self-consistently the neutral H and He density 
distributions in the inner heliosphere (up to $\sim$100 AU), in response to solar gravity, 
radiation pressure and anisotropic ionization processes for the two neutral species. Ionization of H atoms 
is mainly due to their charge-exchange collisions with solar wind (SW) protons and He atoms are mostly ionized by 
solar EUV photons and electron impact. 
We also consider the impact of CX on the SW ion distributions.
This interaction is described in the following reaction:
\begin{equation}
X^{\,Q+} + [H, He] \rightarrow X^{\,*(Q-1)+} + [H^{\,+}, He^+]
\end{equation}

\begin{figure*}[t]
\centering
\epsscale{1.65}
\plotone{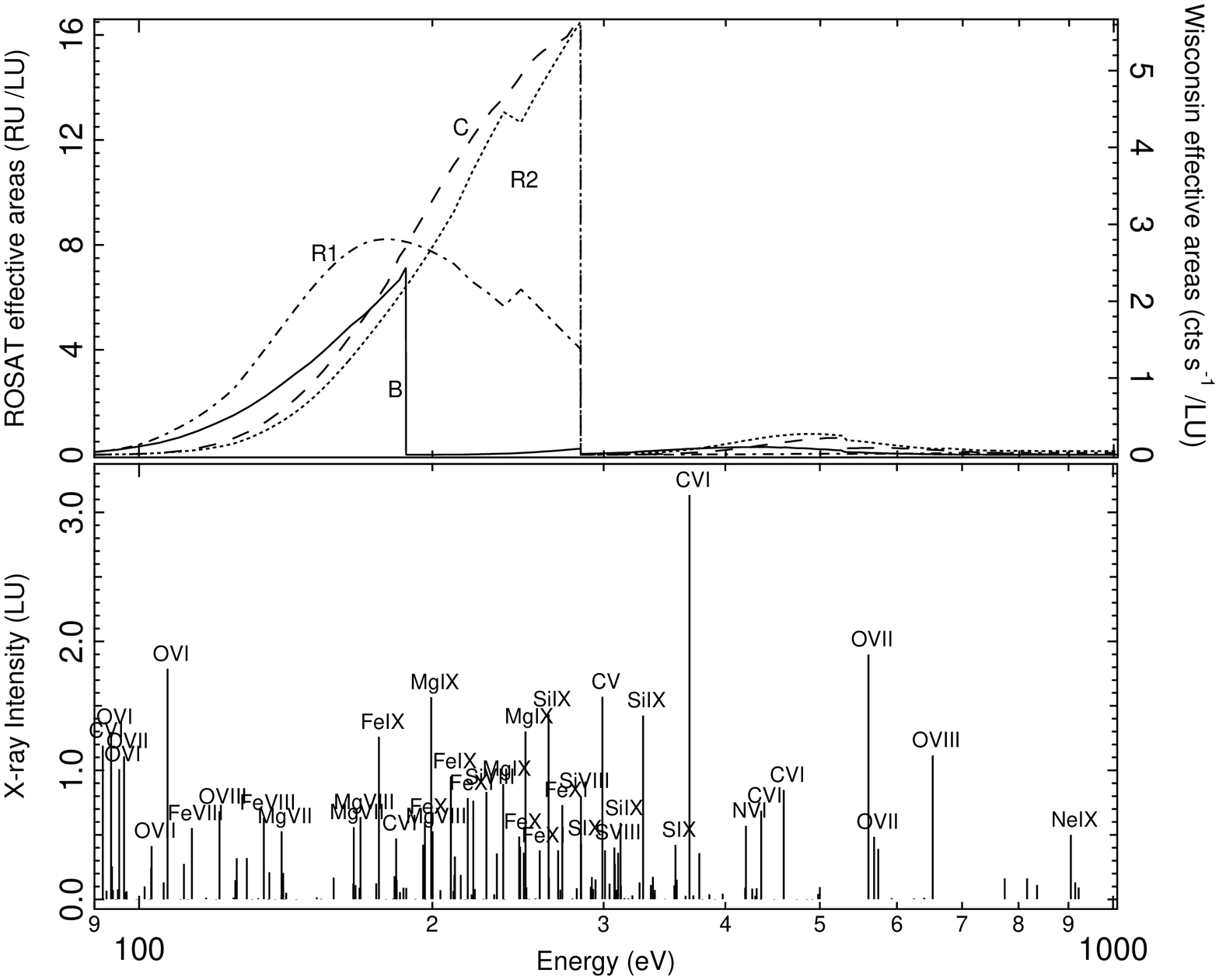}
\caption{\textit{Top panel:} Wisconsin B (plain) and C (dashed), and \rosat\ R1 (dash-dotted) and R2 (dotted) band effective areas. 
\textit{Bottom panel:} Example of calculated SWCX spectra in Line Units (photons cm$^{-2}$ s$^{-1}$ sr$^{-1}$).
The emitting ions are marked above the most prominent lines.}\label{fig:spec1_4}
\end{figure*}

The collision rate per volume unit 
R$_{X^{\,Q+}}$ (cm$^{-3}$ s$^{-1}$) of X$^{\,Q+}$ 
ions with the neutral heliospheric atoms is given
by the equation:
\footnotesize
\begin{eqnarray}
R_{X^{\,Q+}} (r) & = & N_{X^{Q+}}(r)\, \upsilon 
_r\, (\sigma _{(H,X^{\,Q+})}\, n_H(r) +\, \sigma 
_{(He,X^{Q+})}\, n_{He}(r) )\\\nonumber
& = & R_{(X^{\,Q+},H)} (r) + R_{(X^{\,Q+},He)} (r)
\end{eqnarray}
\normalsize
where $\sigma_{(H,X^{\,Q+})}$ and $\sigma_{(He,X^{\,Q+})}$ are the hydrogen and helium CX cross-sections, n$_H$(r) and n$_{He}$(r) 
are the hydrogen and helium density distributions respectively, $\bar \upsilon_r = \bar V_{SW} - 
\bar \upsilon_n \approx \bar V_{SW}$ the relative velocity between SW ions and IS neutrals 
in the inner heliosphere, and $N_{X^{Q+}}(r)$ is the self-consistent solution to the differential equation:
\footnotesize
\begin{eqnarray}
\frac{dN_{X^{\,Q+}}}{dx}\, & =\, & -\, 
N_{X^{\,Q+}}\, (\sigma _{(H,X^{\,Q+})}\, n_H(x) 
+\, \sigma _{(He,X^{\,Q+})}\, 
n_{He}(x))\\\nonumber
& & +\, N_{X^{\,(Q+1)+}}\, (\sigma 
_{(H,X^{\,(Q+1)+})}\, n_H(x) +\, \sigma 
_{(He,X^{\,(Q+1)+})}\, n_{He}(x))
\end{eqnarray}
\normalsize
expressing the evolution of the density distribution of ion X$^{Q+}$ along SW streamlines 
due to production (from CX reactions of ion X$^{(Q+1)+}$) and loss terms.

Cross-section uncertainties are mainly due to instrumental systematic errors and most important 
to the collision energy dependance of cross-sections. Detailed uncertainties for individual ions are 
not given in literature, but average uncertainties of $\sim$30\% at most are reported
\citep[][]{wargelin08}.

Then, we establish emissivity grids in units of (photons cm$^{-3}$ s$^{-1}$):
\begin{equation}
\varepsilon _{i} (r)\,  = 
R_{(X^{Q+},H)}(r)\,Y_{(E_i ,H)} + 
R_{(X^{Q+},He)}(r)\,Y_{(E_i ,He)}
\end{equation}
where Y$_{(E_i ,M)}$ is the photon emission yield (in number of photons) computed for a spectral line of photon energy E$_i$ following CX with 
the corresponding neutral species M (H or He individually). For any line of sight (LOS) and observation date, 
the directional intensity of this spectral line is given by:
\begin{equation} \label{groundI}
\displaystyle I_{E_i}\, (LU) = \frac{1}{4 \pi} \, 
\int_{0}^{\,\sim 100 AU} \varepsilon _{i} (s)\, ds
\end{equation}
which defines the average intensity, in Line Units (LU = photons cm$^{-2}$ s$^{-1}$ sr$^{-1}$), 
of the spectral line for the particular date and LOS, as 
well as the solar cycle phase (minimum or maximum) corresponding at this date. The intensity is somewhat underestimated 
because of the SW ion propagation in the heliosheath up to the heliopause, and in the 
heliotail up to $\sim$3\,000 AU, where all ions are used up. The outer 
heliospheric region is neglected in our model, but estimates yield a 
maximum additional $\sim$20\%\ contribution in the downwind direction, 
with possible effects on the SWCX spectral hardness (see \S \ref{band_ratios}). 

Our original atomic database \citep[][]{khar05} included C$^{5,6+}$, N$^{5,6,7+}$, O$^{6,7,8+}$, 
Ne$^{8,9+}$ and Mg$^{10,11+}$ ions. Exact calculations of the cascading photon spectra were performed 
individually for these ions when they charge exchange with hydrogen and helium respectively. 
Detailed CX collision cross sections taking into account both the neutral target species and 
the solar wind velocity regime were include in the calculations (P. Stancil private communication). 
These calculations have already been used to reproduce observed SWCX spectra from comets with CHIPS 
\citep[][]{sasseen06}. 

The database was recently updated to include Fe$^{7...13+}$, Si$^{5...10+}$, S$^{6...11+}$, Mg$^{4...9+}$ ions 
that emit intense lines in the 0.1-0.3 keV range. Individual emission spectra induced in the charge exchange collisions 
of these ions have very complicated structures because of a large number of intermediate multiplets related 
to different excited states of many-electron ions. The photon yields Y(E,M) for heavier ions were calculated using 
the simplified hydrogenic model \citep{kharchenko00}, which assumes a hydrogenic nature of electronic excited states. In this model, 
the effective charge of hydrogenic ion is computed from an accurate value of the ion ionization potential 
and branching ratios of radiative cascading transitions are chosen to be the same as in all H-like ions. 
Moreover, photon yields were calculated using a single neutral species, 
which means that no distinction between H and He was made. The hydrogenic approximation of the CX emission spectra 
is a quantum mechanical model in which an actual ion spectra may be replaced with the hydrogenic spectra. 
In this model the total energy of emitted photons is defined by an initial state-population and should be 
an accurate quantity matching real spectra. Positions of emission lines do not correspond exactly to real emission spectra, 
but this defect is not very important at the low resolution of the observed spectra. Total cross sections 
of CX collisions for the hydrogenic approximations have been calculated using the over-barrier model \citep{kharchenko01}. 
An example of calculated spectra is presented in figure \ref{fig:spec1_4}-\textit{lower panel}, with the emitting 
ion identifying the most intense lines.

\subsection{Hot Gas thermal emission}
\label{sec:HGmodel}
We use a Raymond-Smith (RS) hot plasma model \citep[][]{raymond77} assuming typical metal abundances 
[He, C, N, O, Ne, Mg, Si, S, Ar, Ca, Fe, Ni] = [10.93, 8.52, 7.96, 8.82, 7.96, 7.52, 7.60, 7.20, 6.90, 6.30, 7.60, 6.30] 
\citep{allen73}. 
We use this model rather than the APEC model \citep[][]{smith01} that superseded it because we are most 
interested in the \oqkev\ range. APEC includes only transitions for which accurate atomic rates are 
available, while the code of Raymond and Smith estimates the emission in the large number of weak lines that 
are known to be present (e.g., from moderately ionized species of Mg, Si, S and Fe) but which 
lack accurate excitation rates and wavelengths. Given the low spectral resolution of the observations 
considered here and our interest in the {\em total} emitted power, the RS model serves very well.

This model gives us X-ray emissivities f$_1$(T) and f$_2$(T) convolved by and summed in the \rosat\ R1 and R2 bands, respectively, 
as a function of temperature such that the total hot gas X-ray flux in these bands is defined as: 
\begin{equation}\label{ILB}
I_{12, LB} = I_{1, LB} + I_{2, LB} = EM(i, T) \cdot (f_1(T) + f_2(T))
\end{equation}
where EM(i, T) is the emission measure for temperature T and look direction $i$. Units of functions f$_1$ and f$_2$ are RU EM$^{-1}$, 
where RU = 10$^{-6}$ cts s$^{-1}$ arcmin$^{-2}$ is the usual \rosat\ detector unit and EM is the typical emission measure unit cm$^{-6}$ pc.

\begin{figure*}[t]
\epsscale{1.66}
\plotone{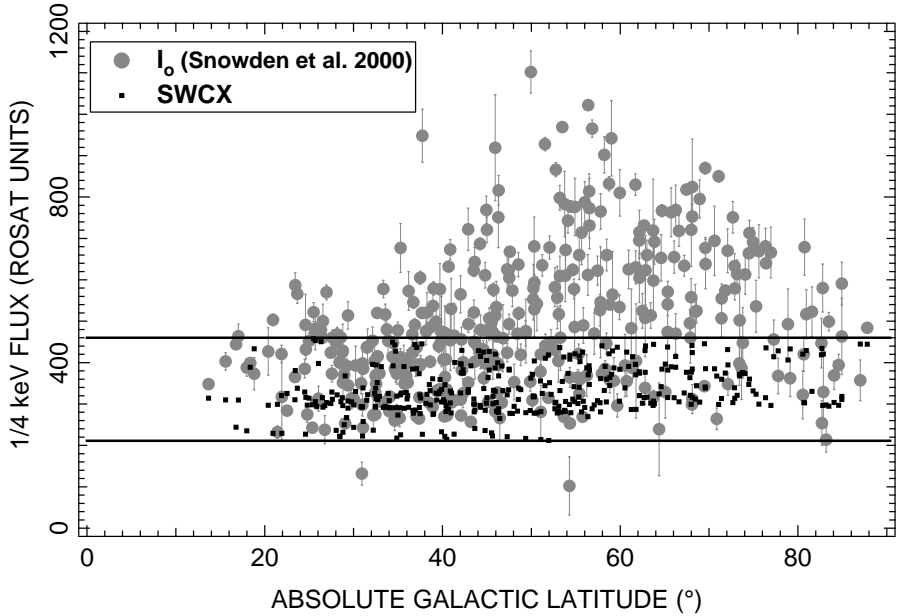}
\caption{Measured unabsorbed (I$_{12, obs}$, from \cite{snowden00}, gray circles) and calculated 
SWCX (black dots) fluxes in the \rosat\ \oqkev\ (R1+R2) band as a function of absolute 
galactic latitude for 378 shadowing observations.
Flux is given in \rosat\ Units (10$^{-6}$ cts s$^{-1}$ arcmin$^{-2}$).}\label{fig:SWCX_Io_lat}
\end{figure*}

Equivalently, the hot gas emission in bands B and C is defined as:
\begin{equation}\label{Ibc}
I_{B\,C, LB} = I_{B, LB} + I_{C, LB} = EM(i, T) \cdot (f_B(T) + f_C(T))
\end{equation}
where f$_B$(T) and f$_C$(T) are the equivalent emissivity functions in the B and C bands derived by the RS plasma code, 
in units of cts s$^{-1}$ EM$^{-1}$.


\section{R1+R2 intensities} \label{CX_IoRosat}

We have calculated SWCX spectra in \rosat\ observation geometry for the shadow field lines 
of sight (LOS) listed in table 1 of \cite{snowden00}, that were observed during the \rosat\ all-sky 
survey. The shadows analysed by \cite{snowden00} were located at high galactic latitudes 
and in general above 15$^{\circ}$ from the galactic plane. The \rosat\ observation geometry is 
defined with the view direction perpendicular to the Sun-satellite direction. Thus, it takes a 
six-month period to build a full-sky map of the soft X-ray intensity. The \rosat\ all-sky 
survey was performed between July 1990 and February 1991, which corresponds to maximum solar 
activity conditions that were taken into account in the SWCX simulations.

Maximum solar activity conditions imply the following input parameters in the SWCX model. We 
consider a radiation pressure to gravity ratio $\mu$ = 1.5 for neutral hydrogen and slightly 
anisotropic ionization rates varying between 8.4$\times$10$^{-7}$ s$^{-1}$ 
at the solar equator and 6.7$\times$10$^{-7}$ s$^{-1}$ at the poles \citep[][]{quem06jgr}. 
For neutral helium, the average lifetime (inverse ionization rate) at 1 AU is 
0.62$\times$10$^7$ s at solar maximum, in agreement with \cite{mcmullin04}. In solar maximum, 
the solar wind is considered to be a complex mix of slow and fast wind states that is in general 
approximated with an average slow wind flux. Slow solar wind flows at $\sim$ 400 km/s and has 
a proton density of $\sim$ 6.5 cm$^{-3}$ at the Earth position. The oxygen content 
with respect to protons is [O/H] = 1/1780. The most important heavy ion charge state abundances 
with respect to oxygen [X$^{q+}$/ O] are: C$^{5,6+}$: 
[0.21, 0.318], O$^{6,7,8+}$: [0.73, 0.2, 0.07], Si$^{8,9,10+}$: [0.057, 0.049, 0.021] 
and Fe$^{8,9,10,11+}$: [0.034, 0.041, 0.031, 0.023] \citep[adopted from ][]{schwadron00}.

\begin{figure*}[t]
\begin{center}
\epsscale{1.65}
\plotone{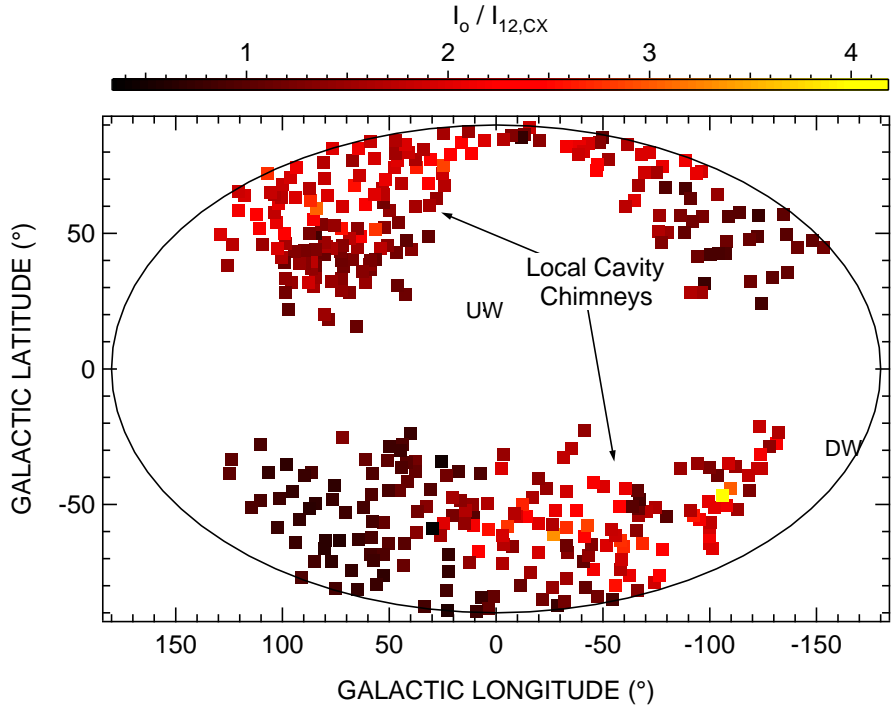}
\caption{Unabsorbed I$_{12, obs}$ R12 flux data over simulated I$_{12,CX}$ SWCX flux ratio map in galactic coordinates 
with an equal-area Aitoff projection. For information we note the UW and DW directions in the sky. 
The LC chimneys appear clearly where the I$_{12, obs}$/I$_{12,CX}$ ratio is the highest.}\label{fig:Io2CX_Aitoff}
\end{center}
\end{figure*}

We have convolved the individual spectra with the \rosat\ R1 and R2 band responses in order to 
calculate the total SWCX flux in these bands, as well as the total R12 (R1+R2) flux. We plot 
the resulting R12 SWCX flux and the unabsorbed I$_{12, obs}$ component from the \cite{snowden00} analysis 
as a function of absolute galactic latitude in figure \ref{fig:SWCX_Io_lat}. I$_{12, obs}$ corresponds 
to the unabsorbed portion of the SXRB that was originally attributed to the LB 
$\sim$10$^6$ K hot gas. X-ray intensities are presented in \rosat\ Units (RU = 10$^{-6}$ cts s$^{-1}$ arcmin$^{-2}$).

The SWCX R12 flux (black dots) varies between 212 and 460 RU with an average value of 332 RU and is 
fairly uniformly distributed across all latitudes. The lower and upper limits calculated 
in the SWCX simulations for average maximum conditions are represented with the plain black horizontal lines.

On the other hand, the unabsorbed I$_{12, obs}$ component (gray circles) derived in the \cite{snowden00} 
analysis has a clear correlation with the absolute galactic latitude. Higher I$_{12, obs}$ 
values are measured towards higher latitudes, where the local cavity is enlarged 
and communicates with the galactic halo through the chimneys.

In the figure it is clear that the SWCX intensity is of the same order as the I$_{12, obs}$ intensity 
measured in low galactic latitudes (up to around 20-25$^{\circ}$). We can conclude, 
then, that the SWCX \oqkev\ flux could account for most of the observed \rosat\ emission 
in the galactic plane. This conclusion implicitly assumes that the highly peaked exospheric SWCX contribution 
has been cleaned from the \rosat\ data, but not the heliospheric contribution.

Fig \ref{fig:Io2CX_Aitoff} is a map in galactic coordinates of the ratio between the unabsorbed \rosat\ emission and 
the computed SWCX contribution. The map clearly reveals the emission from the so-called chimneys 
that connect the Local Cavity to the northern and the southern halo. Our intensity results do not 
preclude that outside these chimneys the totality of the signal is SWCX emission.

\section{Band ratios}\label{band_ratios}

\begin{figure}[t]
\begin{center}
\epsscale{2.1}
\plottwo{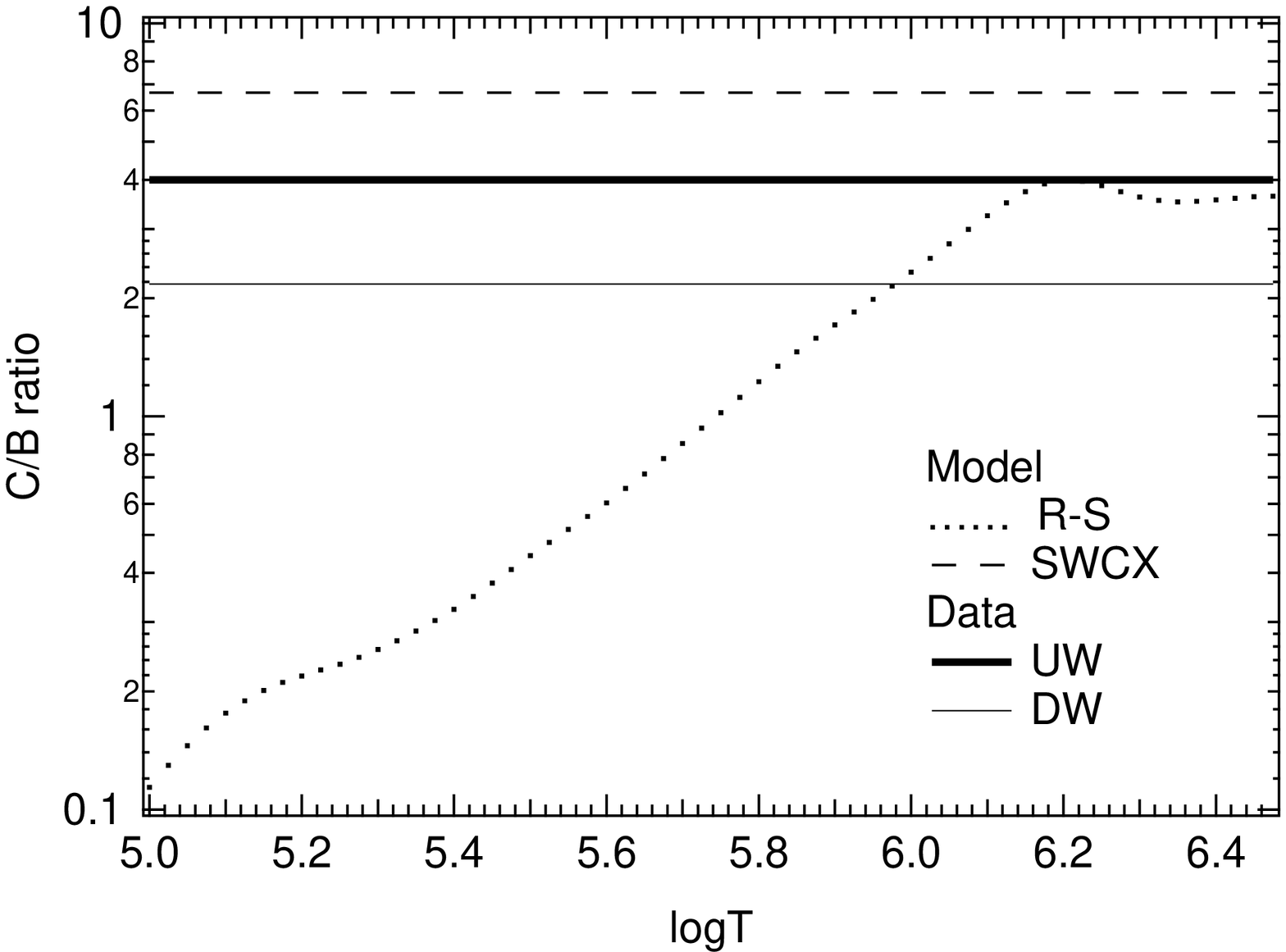}{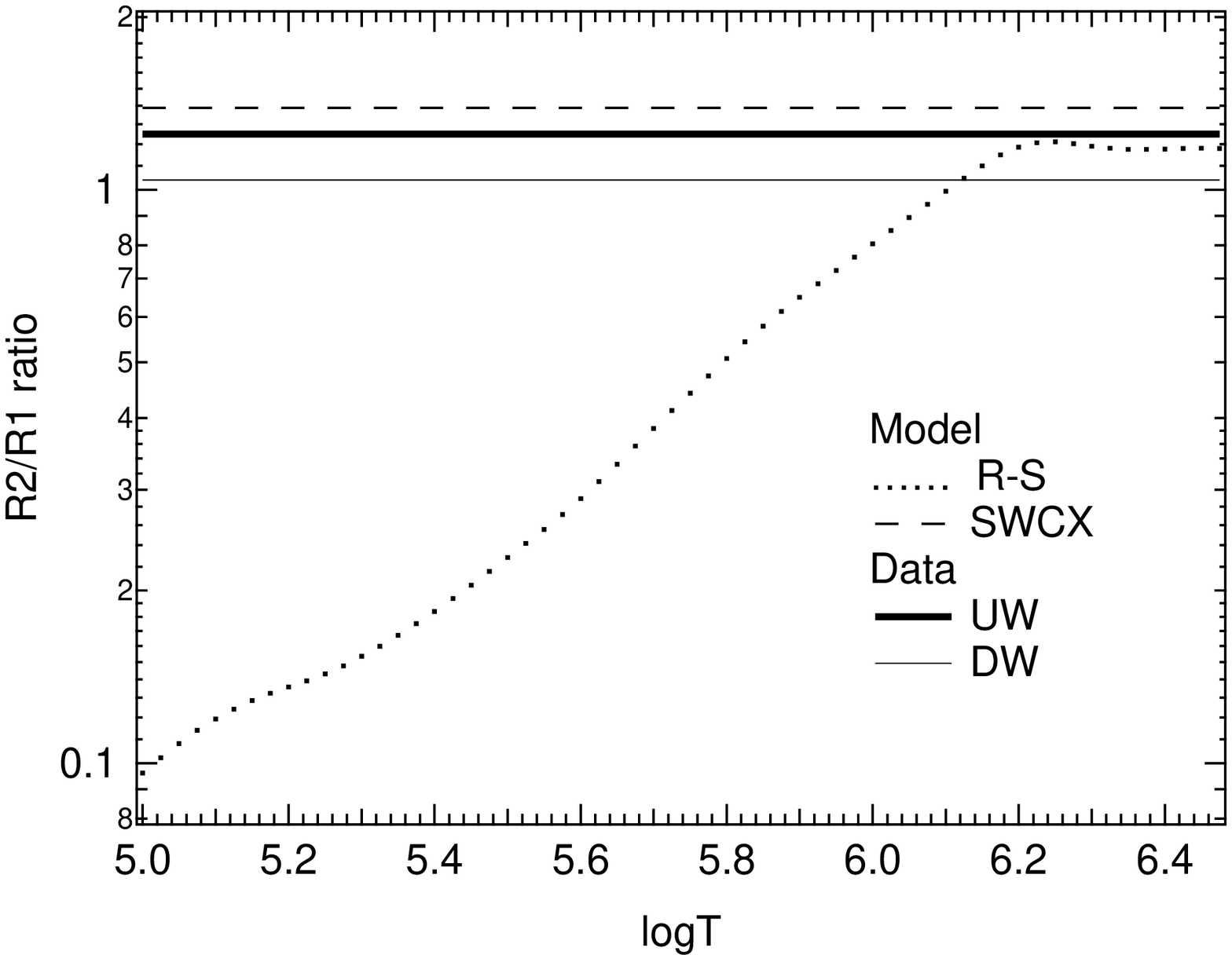}
\caption{Wisconsin (upper panel) and \rosat\ (lower 
panel) band ratios for data (UW-bold, DW-plain 
line), SWCX model (dashed line) and
the R-S hot gas model (dotted line) as a function of temperature.}
\label{fig:ratios}       
\end{center}
\end{figure}

For each SWCX spectrum calculated in the look directions presented in figure 
\ref{fig:SWCX_Io_lat} we have calculated the R2/R1 (\rosat ) and C/B\footnote{The original papers on the Wisconsin 
survey referred to the B/C ratio, but given the extensive use of the R2/R1 ratio in our analysis and 
the rough correspondence between R1 and B, and R2 and C, we refer to the C/B ratio.} (Wisconsin) ratios. We find an average 
R2/R1 (hereafter R$_{CX}$) ratio of 1.39 and an average C/B ratio of 6.67. 
Although both ratios show very little variation across the sky, there is a hardness trend of the 
SWCX spectra with harder spectra towards the downwind direction (UW to DW variations: 
R$_{CX}$ = [1.36 - 1.41], C/B = [6.25 - 7.14]). However, we need to alert the reader that these 
are somewhat uncertain SWCX spectra and therefore somewhat uncertain band ratios. Indeed, as we mentioned in section \ref{sec:model}, exact 
calculations for Fe, Si, Mg, Al are not yet available, and no distinction was done between the neutral targets (H or He), while 
laboratory experiments show that the  energy levels populated after the electron capture and 
the subsequent  radiative cascades may differ significantly for different targets. Since most 
of the SWCX downwind emission is due to the interaction with neutral helium, while on the 
upwind side hydrogen is the main contributor, more precise calculations could have an effect on 
the hardness. Also, although preliminary calculations show an almost negligible effect, 
the heliospheric model cutoff (especially in the DW directions) may be responsible for the 
``loss" of relatively more emission from lower charge states (at relatively lower energies) 
than emission from higher charge states (at relatively higher energies). Thus, the calculated 
SWCX spectra may actually be softer than what is predicted here. It is evident 
that a more detailed calculation taking into account all metals and 
the neutral target nature, as well as detailed cascading collisions (secondary ion production) 
in the outer heliosphere is needed in the future. On the other hand, the interval we find for the ratio can be 
used a reliable value for the average SWCX ratio.

These SWCX ratios have to be compared with the corresponding ratios for thermal emission. The 
latter were obtained as a function of temperature by convolving the Raymond-Smith spectra with the 
\rosat\ band responses R1 and R2 and Wisconsin B and C responses. The results are shown in figure 
\ref{fig:ratios}. Above logT= 6.1 the thermal R2/R1 ratio reaches its maximal value of $\simeq$ 
1.2. It remains however slightly lower than the SWCX ratio of 1.36-1.41. At those temperatures the 
thermal C/B ratio increases to its maximal value of $\simeq$ 4, a value almost half the 
SWCX ratio of 6.24-7.14, i.e. a significant difference. Those curves allow estimates of the 
ratios for combinations of thermal plus SWCX background emissions.


\section{Combination of the heliospheric SWCX and LB hot plasma emission}\label{combination}

For a comparison with the data we consider two regions: one centered on the direction of the 
incoming IS flow at ecliptic coordinates ($\lambda$, $\beta$) $\sim$ (252.3, 8.5)$^{\circ}$ for the IS H flow, 
according to \cite{lall05sci} (upwind -UW- direction) and one looking at the outgoing flow direction (downwind -DW- 
direction). In galactic coordinates the UW direction corresponds to (l, b) = (5.4, 18.9)$^{\circ}$, 
close to the galactic center direction (anti-galactic direction for DW respectively). These two regions 
are also very close to the minimum and maximum values of the hardness ratio derived by 
\cite{snowden90b}, which define the so-called color gradient axis of the soft X-ray background. 
For these two regions we can derive 
average values of observed unabsorbed \oqkev\ emission using the Local Bubble contours in the 
\cite{snowden98} analysis. For the upwind (UW) direction the \rosat\ unabsorbed I$_{12, obs}$ emission we estimate $\sim$325 RU, 
while for the downwind (DW) direction the observed unabsorbed level is found to be 450 RU.

The equivalent B+C intensities in the Wisconsin survey are estimated on average 
$\sim$90~cts s$^{-1}$ and $\sim$125~cts s$^{-1}$ for the UW and DW directions respectively 
\citep{snowden90b}. However, those intensities include both the foreground (assumed LB) 
and more distant components (galactic halo and extragalactic), since the Wisconsin survey 
did not have enough spatial resolution to study the shadowing fields. Moreover, 
the Wisconsin sounding rocket measurements looked in the roughly anti-Sunward direction, 
which should also affect the comparison with the \rosat\ All-Sky Survey in terms of the SWCX 
component spatial distribution. For instance, for the DW look directions, the Wisconsin 
sounding rockets were observing directly through the He cone and should have had a higher 
``contamination" of SWCX emission than the \rosat\ detectors that must have been located 
in crosswind positions on the Earth's orbit in order to observe in the DW directions.

\begin{figure*}
\begin{center}
\epsscale{2.2}
\plottwo{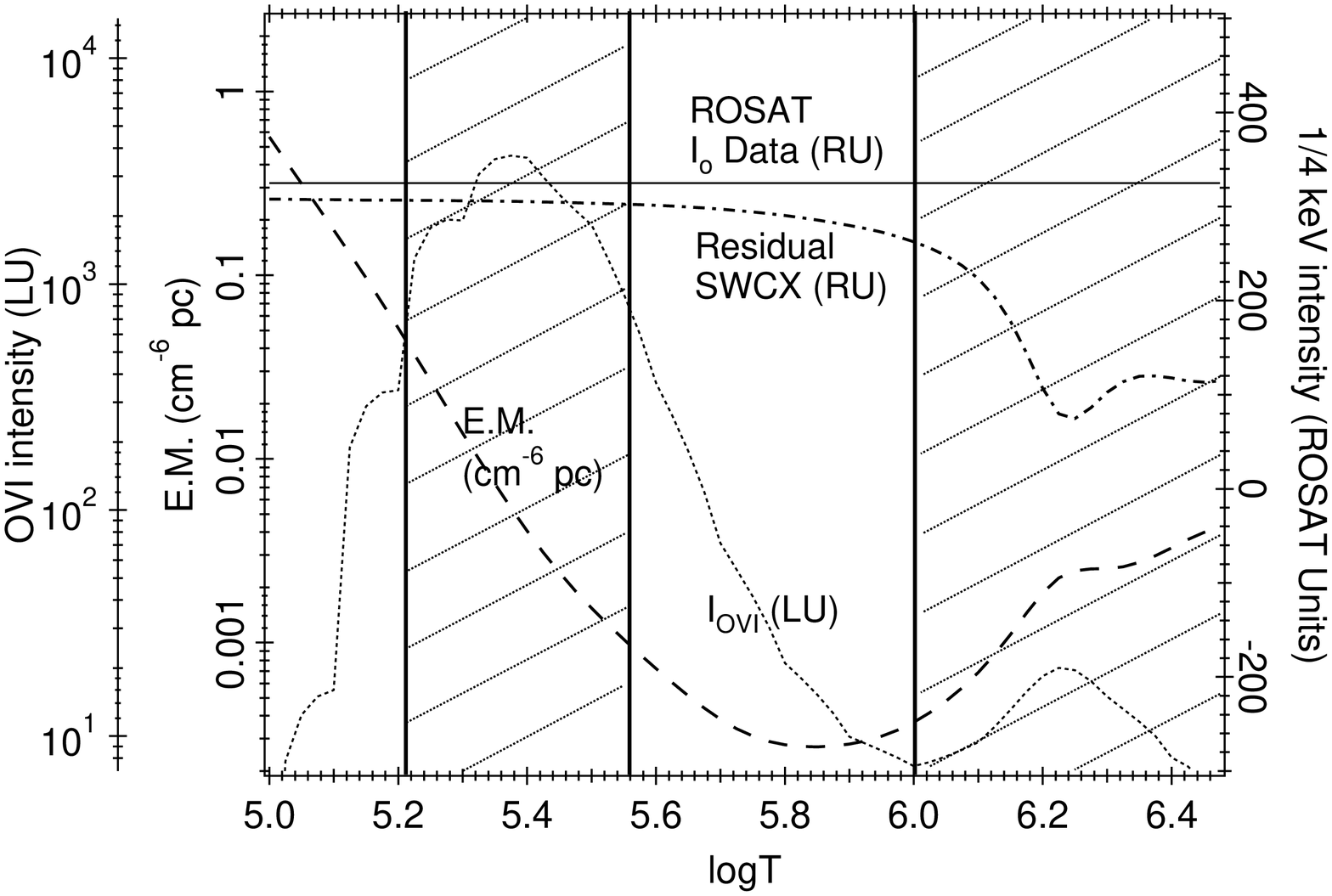}{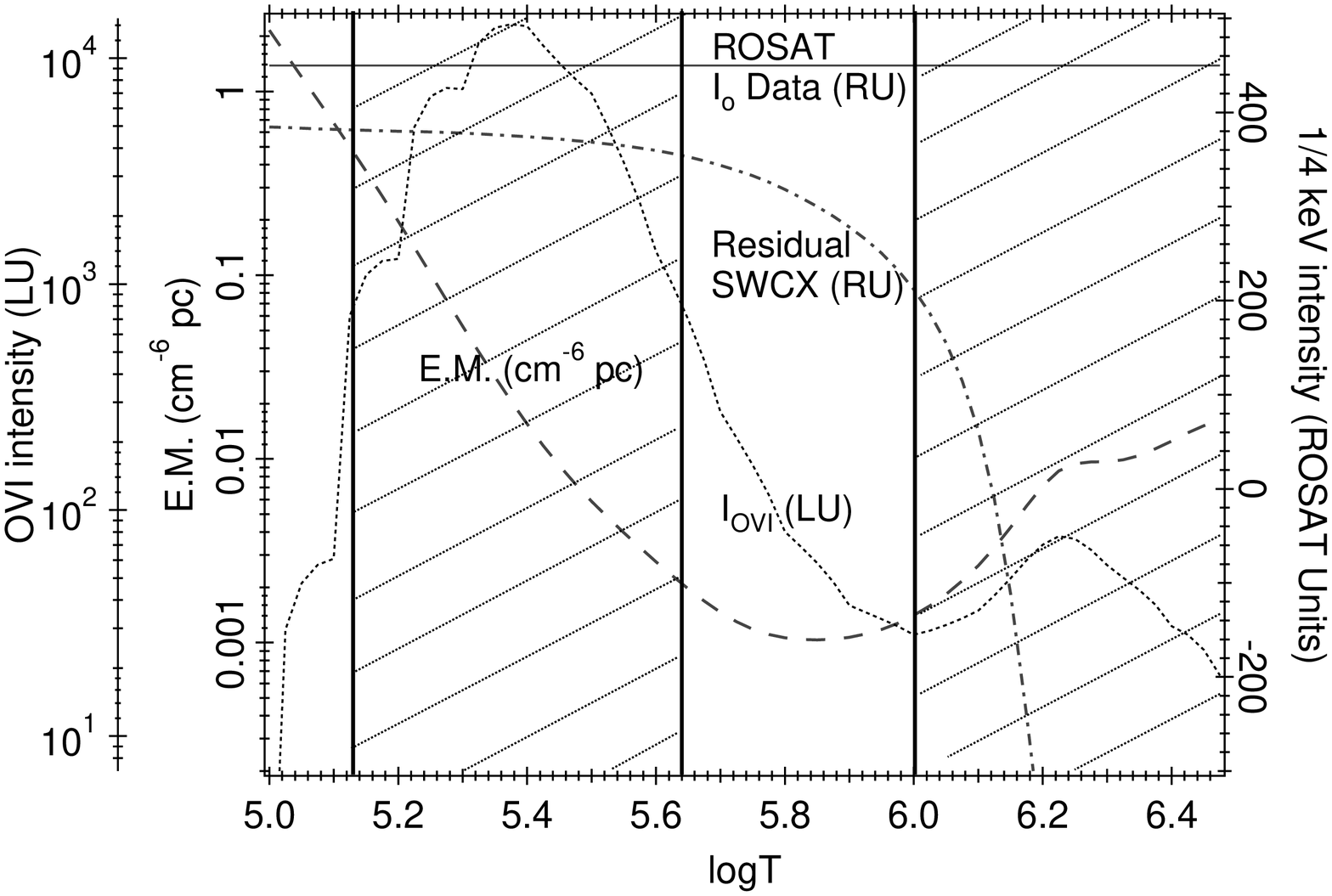}
\caption{Solutions for the EM (dashed line) and residual SWCX emission (dot-dashed line)
as a function of temperature of a hot plasma for observed R12 intensities (I$_{12, obs}$, plain horizontal 
lines) for an upwind (left panel) and downwind (right panel) look directions. Upper 
\ovi\ doublet intensity limit (dotted line) is calculated for the EM values. Discontinuities in the \ovi\ intensity curve 
are due to \ovi\ density interpolation in the Chianti database. The cross-hatched regions show 
the hot gas temperature ranges excluded by \ovi\ intensity \citep[][]{shelton03} and SWCX 
emission constraints.}\label{fig:EM_logT}       %
\end{center}
\end{figure*}

The measured R2/R1 and C/B ratios for these regions are shown superimposed to the models in figure 
\ref{fig:ratios} and listed in Table 1. For the UW area the measured R2/R1 ratio is close to the 
SWCX value, although slightly smaller, while for the DW area it is significantly smaller. For both 
areas the measured C/B ratio is lower than the SWCX ratio. Fig \ref{fig:ratios} shows that for 
both areas a combination of SWCX and thermal emissions may in principle account for the \rosat\ 
measurements, the thermal emission lowering the R2/R1 ratio to achieve the observed value. Similarly, 
independently of the R2/R1 ratio, fig \ref{fig:ratios} also shows that a combination of 
both backgrounds may account for the Wisconsin data, the thermal emission decreasing the C/B ratio to 
achieve the observed value. 

It remains to find a combination satisfying both ratios 
simultaneously. Our attempt to find a solution is the following one. For each assumed temperature 
of the hot gas, we use the R1 and R2 data (and thus the observed ratio) and the SWCX spectral shape to derive the respective 
contributions of SWCX and thermal emission, i.e., we derive which quantity of SWCX induced R12 intensity 
and which emission measure EM for the hot gas lead to the measured intensities and the measured R2/R1 ratio. 
We \textit{a posteriori} calculate the B and C intensities, C/B ratio and the 
\ovi\ emission of the hot gas and compare with the data.

%

The SWCX model predicts a total R12 intensity:
\begin{eqnarray}\label{ICX}
I_{12, CX} = I_{1, CX} + I_{2, CX} & = & I_{1, CX}\cdot (1 + R_{CX})\\
& = & I_{2, CX}\cdot \frac{1 + R_{CX}}{R_{CX}}
\end{eqnarray}
where R$_{CX}$ is the R2/R1 ratio predicted by the SWCX model.

The total unabsorbed flux I$_{12, obs}$(i) measured in the R12 band towards look direction $i$ is the sum of LB hot gas I$_{12, LB}$
and SWCX I$_{12, CX}$ fluxes:
I$_{12, obs}$(i) = I$_{12, CX}$ + I$_{12, LB}$, so that SWCX intensity can be written:
\begin{equation}\label{I12}
I_{12, CX} = I_{12, obs}(i) - I_{12, LB} = I_{12, obs}(i) - EM\cdot (f_1(T)+f_2(T))
\end{equation}
The observed R2/R1 ratio (hereafter R$_{obs}$) towards look direction $i$ is defined by the equation:
\begin{equation}\label{Ro}
R_{obs}(i) = \frac{I_{2, CX} + I_{2, LB}}{I_{1, CX} + I_{1, LB}}
\end{equation}

Resolving equation \ref{Ro} by using equations \ref{ILB} to \ref{I12}, we find the hot gas emission measure EM(i, T) 
as a function of temperature, total R12 measured intensity I$_{12, obs}$(i) and measured R$_{obs}$(i) ratio towards look direction $i$.
\begin{equation}\label{EM}
EM(i, T)\,= 
\frac{R_{CX}\,-\,R_{obs}(i)}{(R_{obs}(i)\,+\,1)\, \cdot 
\, (R_{CX}\cdot f_1(T)\, -\, f_2(T)\,)}\cdot I_{12, obs} 
(i)
\end{equation}
We calculate the emission measure for the two UW and DW directions defined above using the following numerical values: 
(i) the R$_{CX}$ ratio is constant and equal to 1.39, (ii) observed values of the unabsorbed portion 
of the \oqkev\ emission in the R12 band are I$_{12, obs}$(UW, DW) = (325, 450) RU as derived from the LB contours in the 
\cite{snowden98} analysis for the UW and DW (respectively galactic and anti-galactic) directions, and 
(iii) the corresponding observed R2/R1 ratio is R$_{obs}$ = 1.25 and 1.04 for the UW and DW directions respectively.

\setlength{\extrarowheight}{0.5mm}
\begin{table*}[t]
\begin{center}
\caption{Temperature (logT) limits and E.M. solutions for upwind (UW) and downwind (DW) look directions 
when combining a SWCX and RS hot plasma code. In the lower part of the table we include the observational input considered.\label{tbl-sum}}
\begin{tabular}{ccccccccccccc}
\tableline\tableline
&\multicolumn{5}{c}{Local Bubble} &  & \multicolumn{3}{c}{SWCX} 
	& & \multicolumn{2}{c}{(LB + SWCX)\tablenotemark{c}}\\\cline{2-6}\cline{8-10}\cline{12-13}
Look & logT & E\,M & I$_{12,LB}$ & I$_{B}$\tablenotemark{a} & I$_{C}$\tablenotemark{a} 
	& &I$_{12,CX}$\tablenotemark{b}  & I$_B$ & I$_C$ & &I$_{BC}$ & C/B\\
Direction & & (10$^{-4}$ cm$^{-6}$ pc) & (RU) &\multicolumn{2}{c}{(cts s$^{-1}$)} 
	& &(RU) & \multicolumn{2}{c}{(cts s$^{-1}$)}& &(cts s$^{-1}$) & \\
\tableline
UW & 5.64 & 5.5 & 25 & 3.8  & 2.6 & & 300 & 10.  & 66.5 & &  82.9 & 5.\\
DW & 	  & 21. & 96 & 14.5 & 10. & & 354 & 11.8 & 78.8 & & 115.1 & 3.37\\

\multicolumn{13}{c}{} \\
UW & 6.00 & 3.7 & 62  &  4.8 & 11.1 & & 263 & 9.2 & 58.  & &  82.7 & 5.\\
DW & 	  & 14.1 & 238 & 18.3 & 42.5 & & 212 & 7.7 & 46.5 & & 114.1 & 3.57\\

\tableline
& \multicolumn{4}{c}{Observational Input} & \\\cline{2-5}
&I$_{12, obs}$(RU) & R2/R1 (R$_{obs}$) & I$_{BC}$ (cts s$^{-1}$) & C/B & \\
UW & 325 & 1.25 & $\sim$ 90 & 4. & \\
DW & 450 & 1.04 & $\sim$ 125 & 2.17 & \\
\tableline
\end{tabular}
\tablenotetext{a}{I$_{B}$\,=\,EM\,$\cdot$\,f$_B$(T), I$_{C}$\,=\,EM\,$\cdot$\,f$_C$(T)}
\tablenotetext{b}{I$_{12,CX}$\,=\,I$_{12, obs}$ - I$_{12,LB}$}
\tablenotetext{c}{I$_{BC}$\,=\,I$_B$(LB+CX)\,+\,I$_C$(LB+CX), C/B\,=\,I$_B$(LB+CX)\,/\,I$_C$(LB+CX)}
\end{center}
\end{table*}

In figure \ref{fig:EM_logT} we show the resulting EM and the portion of the total emission due to the SWCX mechanism 
(called residual emission) as derived from equations \ref{EM} and \ref{I12} respectively as a function of logT 
and for the two look lines. We show superimposed the R12 measured intensities in those directions. 
For the calculated EM and corresponding temperatures we have also added to figure \ref{fig:EM_logT} the 
intensity of the \ovi\ doublet at $\bar{\lambda}$ = 1034~\AA\ (1032~\AA\ and 1038~\AA ). 
In order to calculate this \ovi\ doublet emission we have used equation (5) of \cite{shull94} 
and assumed that interstellar O abundance is 8.5$\times$ 10$^{-4}$. We also assume that the \ovi\ ion proportion depends 
on temperature according to the Chianti database formulae for collisional equilibrium \citep{landi06}.

In order to delimit the possible temperature solutions for the LB hot gas, we place the following constraints: 
i) We assume that the SWCX model is accurate enough to ensure that the heliospheric emission in the R12 band 
cannot be lower than $\sim$ 212 RU (lower limit in figure \ref{fig:SWCX_Io_lat}). This gives us 
(from the right panel of Fig.\ref{fig:EM_logT}) an upper 
limit of logT = 6 in temperature. ii) We use the observed upper limit of \ovi\ doublet intensity, 
reported at $\sim$ 800 LU \citep[][]{shelton03}, which gives us two temperature limits at logT 
= 5.13 and 5.64 (extreme limits in the DW direction). The interval between those two temperatures is forbidden because 
the corresponding \ovi\ column densities (and thus the \ovi\ intensity) are too high to match observations. 
Temperatures below 10$^{5.13}$ K would predict extremely strong \cvi\ and \nv\ absorption toward nearby stars, 
which have not been observed \citep[e.g.][]{lehner03, welsh05} so we do not consider it a realistic solution. 
The limits of valid temperature intervals are marked by the vertical bold lines and the cross-hatched 
regions in figure \ref{fig:EM_logT} show the excluded temperature ranges. The two most plausible hot gas temperature limits (logT = 5.64, 6.0) 
along with the corresponding UW and DW emission measures and residual SWCX emission are summarized in table \ref{tbl-sum}.

We also calculate for the two boundary solutions the corresponding SWCX intensities and the thermal emission intensities 
in the B and C bands by convolving our simulated SWCX spectra and the hot gas spectra with the band responses. 
For the two temperatures the total hot gas and SWCX intensity in (B + C) band is found to be about 
83 cts s$^{-1}$ and about 115 cts s$^{-1}$ for the UW and DW directions respectively. This similarity arises from the similarity 
between the wavelength intervals covered by the B and C bands and the R1 and R2 bands (see fig. \ref{fig:spec1_4}). The C/B ratio 
derived from this analysis is 5.0 and $\sim$3.5 for the UW and DW directions accordingly, for temperatures above 10$^{5.64}$ K.

The (B+C) total intensity is consistent with the lower values reported in the Wisconsin survey 
\citep[][]{snowden90b}, which correspond to the lower galactic latitudes. 
Moreover, \cite{snowden90b} did not proceed with a shadowing analysis of the Wisconsin data, so the reported values 
include both local and more distant absorbed components and are expected to be higher than the hot gas 
and SWCX combination we present here.

However, the C/B ratio computed in the analysis (5 to $\sim$3.6 from UW to DW, depending on temperature) is inconsistent with 
the observed value, especially in the DW direction (observed $\sim$2.2), suggesting that we should 
need more hot gas emitting in the B band. This inconsistency cannot be attributed to 
the absorbed portion of emission included in the Wisconsin data analysis because the absorbed 
component is a high-T gas giving a harder spectrum since absorption is more effective in lower energies.

This inconsistency of the DW C/B ratio is important, since it seems difficult to explain in the context of our study. 
As a matter of fact, as can be seen in table \ref{tbl-sum}, the SWCX contribution in the C band is large 
whatever the temperature, and reaching a C/B ratio of 2.2 requires a very small SWCX emission, 
in our sense far from realistic. Again, as we discussed in the introduction and it was shown in the \cite{bellm05} study, the B 
intensity is higher than expected from the models. This seems to remain true (and even worse) 
when taking into account the SWCX contribution.

\section{Discussion and conclusions}\label{discussion}

We have modeled the intensity and spectral characteristics of the heliospheric SWCX emission 
at the time of the \rosat\ survey and compared with the unabsorbed, local emission derived by 
\cite{snowden00} in the \oqkev\ band. The results show that the SWCX emission can account for most of the 
total intensity recorded in the R1+R2 bands for most of low latitude lines-of-sight. A map of the 
heliospheric SWCX portion of the total signal clearly reveals the high latitude chimneys to the 
halo as the only regions unambiguously dominated by hot gas emission. Such a result can be interpreted 
as meaning that little or no hot gas exists within the galactic disk.

The spectral characteristics however reveal more complexity and preclude such a simple scenario. 
The SWCX band ratios disagree with the observations, especially towards the galactic anti-center 
and at low energies (C/B). We have thus searched for a combination of SWCX 
and thermal emission from hot gas in equilibrium and solar abundances able to reproduce the data. 
Our study shows that a combination of SWCX and thermal emission can reproduce the data in the 
galactic center hemisphere at low latitudes. For this solution the SWCX emission strongly dominates. 
The temperature of the hot gas is constrained within the interval 10$^{5.64}$-10$^{6}$. The 
upper limit is constrained by the lower limit on the SWCX intensity. This upper limit can be 
considered as firmly determined, thanks to recent observational studies above 0.3 keV that have 
confirmed the validity of our model \citep[][]{kout07}. The temperatures lower than 10$^{5.64}$ are excluded by \ovi\ 
emission observations \citep{shelton03} and interstellar ion absorption lines toward nearby stars.

On the other hand, it is difficult to fit with such a combination the Wisconsin data. In the UW (galactic center direction) a combination 
of hot gas and SWCX emission gives (B+C) intensities as well as C/B ratios roughly compatible with the observed values 
for several different temperature ranges. The main difficulty is the impossibility to account 
for the very low C/B ratio measured towards the galactic anti-center direction with the present input models used in our study. 
We note that the high B intensity is also clearly a problem for any hot gas solution, including the very high depletion hypothesis, as 
shown by the study of \cite{bellm05}. The SWCX contribution, which hardens the spectra, reinforces this difficulty. 

However, further investigation is required on the model's uncertainties in order 
to quantify their influence on the spectral hardness of SWCX emission. Further analysis 
of the hydrogenic ion approximation is needed, since hydrogenic ions tend to emit photons 
at higher energies following CX (since high-n to ground transitions are generally allowed) 
than the multi-electron ions considered here (because selection rules and more complicated 
atomic structure lead to more cascades before the final transition to ground. Therefore, 
it is likely that the model spectra have significantly 
less flux at low energies than they should. Given the steeply decreasing effective areas 
at lower energies (see Fig. \ref{fig:spec1_4}), this would help explain some of 
the discrepancies seen in the C/B ratio. Also, the fact that no distinction 
was made between the neutral targets (H or He) in the \oqkev\ calculations, would also have 
an effect on the spectral hardness, since there is an effect on the electron capture level 
(roughly proportional to $\sqrt{13.6\,eV/I_n}$, where $I_n$ is the neutral target ionization 
potential), the capture level with He being somewhat lower than with H. The \rosat\ observation 
geometry tends to smooth out those differences, but the SWCX spectrum will tend to be softer 
in the DW directions (i.e., looking through or near the He cone), again helping to explain 
some of the C/B ratio anomaly. Finally, at the low energies of the B band, more detailed calculations 
of the heliospheric and magnetospheric signals must be performed, especially 
the low energy secondary SWCX emissions in the heliosheath and heliotail, 
i.e. subsequent recombinations of partially neutralized solar wind high ions.

\acknowledgments

The authors wish to thank Dan McCammon and Steve Snowden for comments 
and useful discussions on the \rosat\ and Wisconsin surveys 
and band responses. The work presented has benefited from very useful discussions 
at the ISSI workshop `From the Outer Heliosphere to the Local Bubble' and at 
the `Local Bubble and Beyond II' meeting. \textbf{We are extremely greatful 
to the anonymous referee for his attentive report and valuable comments which 
resulted in significantly improving the paper.}





\end{document}